\documentstyle[prl,aps,floats,epsf,color]{revtex}
\begin{document}
\baselineskip=12pt
\def\be{\begin{equation}}
\def\ee{\end{equation}}
\def\bea{\begin{eqnarray}}
\def\eea{\end{eqnarray}}
\def\E{{\rm e}}
\def\bearst{\begin{eqnarray*}}
\def\eearst{\end{eqnarray*}}
\def\peleven{\parbox{11cm}}
\def\peffec{\peight{\bearst\eearst}\hfill\peleven}
\def\pspace{\peight{\bearst\eearst}\hfill}
\def\ptwelve{\parbox{12cm}}
\def\peight{\parbox{8mm}}
\twocolumn[\hsize\textwidth\columnwidth\hsize\csname@twocolumnfalse\endcsname

\title
{Observational Constraints on a Variable Dark Energy Model}

\author{M. Sadegh Movahed$^{1,2,3}$, Sohrab Rahvar$^{1,2}$}
\address{$^{1}$Department of Physics, Sharif University of
Technology, P.O.Box 11365--9161, Tehran, Iran}
\address{$^{2}$ Institute for Studies in theoretical Physics and Mathematics, P.O.Box 19395-5531, Tehran, Iran}
\address{$^{3}$ Iran Space Agency, P.O.Box 199799-4313, Tehran, Iran}

\vskip 1cm

 \maketitle


\begin{abstract}

 We study the effect of a phenomenological parameterized quintessence
model on low, intermediate and high redshift observations. At low
and intermediate redshifts, we use the Gold sample of supernova Type
Ia (SNIa) data and recently observed size of baryonic acoustic peak
from Sloan Digital Sky Survey (SDSS), to put constraint on the
parameters of the quintessence model. At the high redshift, the same
fitting procedure is done using WAMP data, comparing the location of
acoustic peak with that obtain from the dark energy model. As a
complementary analysis in a flat universe, we combine the results
from the SNIa, CMB and SDSS. The best fit values for the model
parameters are $\Omega_m = 0.27^{+0.02}_{-0.02}$ (the present matter
content) and $w_0=-1.45^{+0.35}_{-0.60}$ (dark energy equation of
state). Finally we calculate the age of universe in this model and
compare it with
the age of old stars and high redshift objects.
\end{abstract}
\hspace{.3in}
\newpage
 ]
\section{Introduction}
Observations of the apparent luminosity and redshift of type Ia
supernovas (SNIa) provided us with the main evidence for the
accelerating expansion of the Universe \cite{ris,permul}. A
combined analysis of SNIa and the Cosmic Microwave Background
radiation (CMB) observations indicates that the dark energy fills
about $2/3$ of the total energy of the Universe and the remaining
part is the dark matter and a few percent in the Baryonic form
\cite{bennett,peri,spe03}.

The "cosmological constant" which was introduced by Einstein to
have a static universe, can be a possible solution for the
acceleration of the universe \cite{wein}. The cosmological
constant in Einstein field equation is a geometrical term,
however, it can be regarded as a fluid with the equation of state
of $w=-1$. There are two problems with this scenario, namely the
two dark energy problems of the {\it fine-tuning} and the {\it
cosmic coincidence}. Within the framework of quantum field theory,
the vacuum expectation value of the energy momentum tensor
diverges as $k^4$. A cutoff at the Planck energy leads to a
cosmological constant with $123$ orders of magnitude larger than
the observed value of $10^{-47}$ GeV$^{4}$. The absence of a
fundamental mechanism which sets the cosmological constant to zero
or to a very small value is the cosmological constant problem. The
second problem, cosmic coincidence, states that since the energy
densities of dark energy and dark matter scale so differently
during the expansion of the Universe, why are they nearly equal
today?

One of the solutions to this problem is a model with decaying
cosmological constant from the Planckian area at the early
Universe to a small enough at the present time. Dolgov (1983)
proposed a massless non-minimally coupled scalar field to the
gravity with a negative coupling constant to solve this problem
\cite{dol83}. However, this model provides a time varying
Gravitational constant which strongly contradicts the upper limits
from the Viking radar range \cite{hel83} and lunar laser ranging
experiments \cite{wil96}. A non-dissipative minimally coupled
scalar field, so-called Quintessence model can also play the role
of time varying cosmological constant \cite{wet88,amen,peb88}. The
ratio of energy density of this field to the matter density
increases by the expansion of the universe and after a while the
dark energy becomes the dominated term in the energy-momentum
tensor. Tuning the parameters of this model it can produce the
value of $\Lambda$-term both for the early universe and the
present time.

One of the features of the quintessence model is the evolution of
the equation of state of dark energy during the expansion of the
Universe. Various models depending on the potential for the scalar
field as k-essence \cite{arm00}, Tachyonic matter \cite{pad03},
Phantom \cite{cal02,cal03} and Chaplygin gas \cite{kam01} provide
different time dependent functions for the equation of state
{\cite{cal03,arb05,wan00,per99,pag03,dor01,dor02,dor04}. There are
also phenomenological models, parameterize the equation of state
of dark energy in terms of redshift \cite{che01,lin03,sel04}.
Here we examine a simple phenomenological parameterization for
the variable dark energy, proposed by Wetterich (2004)
\cite{w04}. In this parameterization the variable dark energy is
expressed in terms of three parameters. The first two parameters
are the dark energy density, $\Omega_{\lambda}$ and dark energy
equation of state $w_0$ at the present time. The third parameter
$b$ is the bending parameter which can be expressed in terms of
the fraction of dark energy at early universe. The equation of
state of this model depends on $b$ and $w_0$ as:
\begin{equation}
w(z;b,w_0)=\frac{w_0}{\left[1+b\ln(1+z)\right]^2}, \label{eq1}
\end{equation}
Using the continuity equation, the density of dark energy changes
with the redshift as:
\begin{equation}
\rho_{\lambda}(z;b,w_0)=\rho_\lambda(1+z)^{3[1+\bar{w}(z;b,w_0)]},
\end{equation}
where $\bar{w}(z;b,w_0)={w_0}/\left[1+b\ln(1+z)\right]$. In this
paper we compare this model in a flat universe with the cosmological
observations such as SNIa, CMB shift parameter and  Large Scale
Structure (LSS) and constrain the parameters of the model.

The organization of the paper is as follows: In
Sec.~\ref{lowredshift} we study the effect of the parameters of the
dark energy model on the age of Universe, comoving distance,
comoving volume element and the variation of angular size by the
redshift \cite{alc79}. In Sec.~\ref{sn} we use the Gold sample of
Supernova Type Ia data \cite{R04} to constrain the parameters of the
model. In Sec. \ref{cmb} the position of the observed acoustic
angular scale on CMB is compared with that of quintessence model. In
Sec. \ref{comb} with the combined SNIa$+$ CMB$+$ SDSS data we put
better constraints on the parameters of this model. Finally the age
of the universe determined by this model is compared with the age of
the old stars and old high redshift objects as the consistency test.
The conclusions are given in Sec.~\ref{conc}.

\section{the effect of variable dark energy on the geometrical parameters}
\label{lowredshift} In this section we study the geometrical
effect of the dark energy on the observable parameters of the
universe as itemized as follows:

\begin{itemize}
\item
{\it comoving distance:} The radial comoving distance of an object
located at a given redshift $z$ is one of the basis parameters of
the cosmology. Using the null geodesics in the FRW metric, the
comoving distance can be obtained by:
\begin{equation}
r(z;b,w_0)  =  \int_0^z {dz'\over H(z';b,w_0)}, \label{comoving}
\end{equation}
where $H(z;b,w_0)$ is the Hubble parameter and for the redshifts
smaller than the radiation dominant epoch $(z<z_{eq})$, it can be
expressed in terms of Hubble parameter at the present time, $H_0$,
the matter and dark energy content of the universe as:
\begin{equation}
H^2(z;b,w_0)=H_0^2[\Omega_m(1+z)^3+\Omega_\lambda(1+z)^{3[1+\bar{w}(z)]}],
\label{eq4}
\end{equation}

\begin{figure}
\epsfxsize=9.5truecm\epsfbox{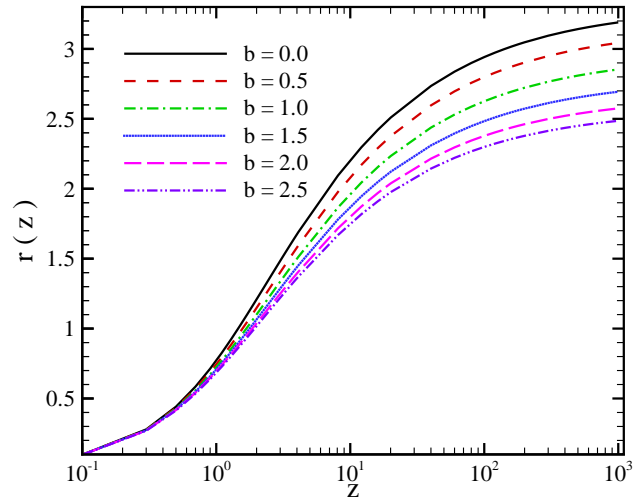} \narrowtext
\caption{Comoving distance, $r(z;b,w_0)$ (in unit of $c/H_0$) as a
function of redshift for various values of bending parameters. }
\label{fig:rz}
 \end{figure}

By numerical integration of equation (\ref{comoving}), the
comoving distance as a function of redshift can be obtained. The
dependence of comoving distance in terms of redshift for various
bending factors is shown in Figure~\ref{fig:rz}. This diagram
shows that the comoving distance is more sensitive to $b$ at
higher redshifts than lower redshifts. Increasing the bending
parameter leads the dark energy dominance at the higher redshifts
which results a slow growth of $r(z;b,w_0)$. For redshifts around
$z=0$ we can expand equation (\ref{comoving}) as:
\begin{equation}
r(z)=H_0^{-1}\left[ z -{3\over 4}z^2(1+ \Omega_{\lambda} w_0) +
\frac{\Omega_{\lambda}w_0}{2}b z^3 + \cdots \right].
\label{eq:hubblelaw}
\end{equation}


One of the main applications of the comoving distance is in the
luminosity distance calculation. In the next section we use SNIa
data as the standard candle in the cosmological scales to confine
the parameters of the dark energy model.

\item
{\it angular size:} The apparent angular size of objects at the
cosmological scales is another observable that can be affected by
the dark energy model. If $D$ is the physical size of an object that
subtends an angle $\theta$ to the observer, for small $\theta$ we
have:
\begin{equation}
D=d_A \theta \label{as}
\end{equation}
where $d_A=r(z;b,w_0)/(1+z)$ is called the angular diameter
distance. One of the main applications of equation (\ref{as}) is
on measurement of matter content of the universe by observing the
apparent angular size of acoustic peak on CMB map and baryonic
acoustic peak at lower redshifts. A variable dark energy can
change the comoving distance to the observer and consequently the
apparent size of the acoustic peak. So measurement of the angular
size of objects in various redshifts (so-called Alcock-Paczynski
test) can probe the variable dark energy \cite{alc79}. The
variation of apparent angular size $\Delta\theta$ in terms
variations of redshift $\Delta z$ can be determined as:
\begin{equation}
{\Delta z\over \Delta \theta} = \frac{H(z;b,w_0)r(z;b,w_0)}{\theta}
\label{alpa}
\end{equation}

\begin{figure}
\epsfxsize=9.5truecm\epsfbox{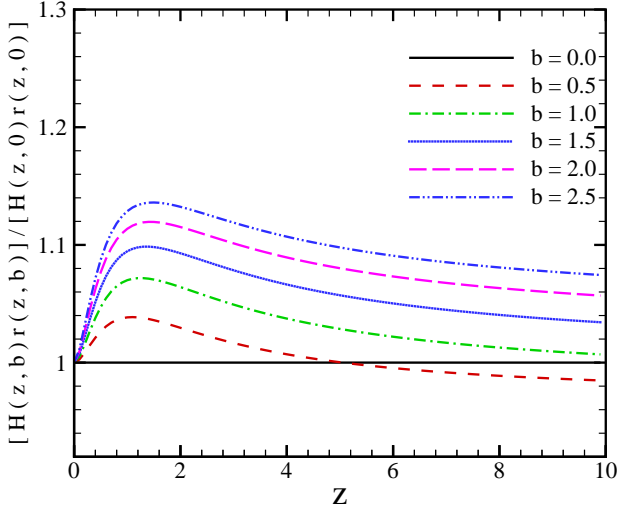} \narrowtext
\caption{Alcock-Paczynski test, compares $\Delta z/{\Delta
\theta}$ normalized to the case of $\Lambda$CDM model as a
function of redshift. } \label{fig:hr}
 \end{figure}
Figure~\ref{fig:hr} shows $\Delta z/ \Delta \theta$ in terms of
redshift, normalized to the case with $b=0$ ($\Lambda$CDM model).
According to Figure~\ref{fig:hr}, the variation of the apparent
angular size is sensitive to the bending parameter at higher
redshifts. The advantage of Alcock-Paczynski test is that we need
standard ruler in the universe instead of the standard candle.
Recently a possible Alcock-Paczynski test for measuring the dark
energy parameters has been proposed by high-z observation of the
power spectrum of large scale structures at 21cm wavelength.
Observation of 21cm fluctuations will enable us to determine the
angular diameter and the Hubble constant \cite{nus04,bar05}. The
Ly-$\alpha$ forest of close QSO pairs may also measure
$\Omega_\lambda$ and its variation with time \cite{eri03}.
\item{\it comoving volume element:} The other geometrical parameter is the comoving
volume element which is the basis of number-count tests, such as
lensed quasars, galaxies, or clusters of galaxies. The comoving
volume element in terms of comoving distance and Hubble parameters
is given by:
\begin{equation}
f(z;b,w_0) \equiv {dV\over dz d\Omega} = r^2(z;b,w_0)/H(z;b,w_0).
\end{equation}
As shown in Figure \ref{fig:v}, the comoving volume element
reaches to its maximum value around $z\simeq2$ and for larger
bending parameters the position of the peak shifts to the smaller
redshifts.

\begin{figure}
\epsfxsize=9.75truecm\epsfbox{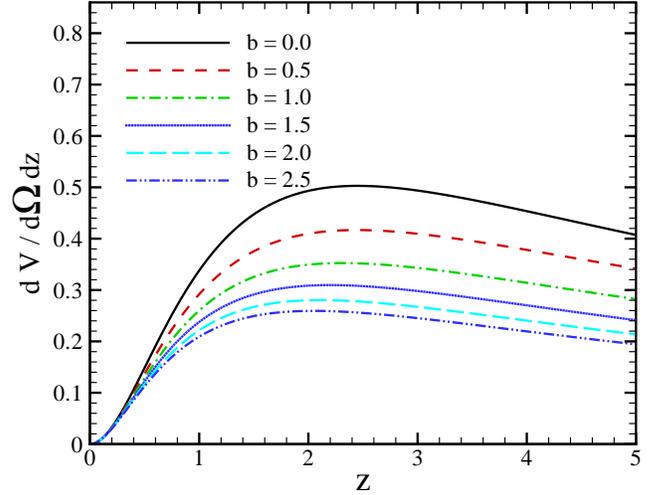} \narrowtext \caption{The
comoving volume element in terms of redshift for various bending
parameters. Increasing the bending parameters makes the position
of maximum value of volume element shifts to the smaller
redshifts. } \label{fig:v}
 \end{figure}

\item{\it age of the universe:}
The age of universe is another observable parameter that can be
used to constrain the parameters of dark energy models. Studies on
the old stars \cite{carretta00} suggests an age of $13^{+4}_{-2}$
Gyr for the universe. Richer et. al. \cite{richer02} and Hansen
et. al. \cite{hansen02} also proposed an age of $12.7\pm0.7$ Gyr,
by using the white dwarf cooling sequence method. For a detailed
review on the cosmic age see \cite{spe03}. The universe age in
varying dark energy models depends on the parameters of the model.
Integrating the age of universe from the beginning of universe up
to now, we obtain the age of universe as follows:
\begin{equation}\label{age}
t_0(b,w_0) = \int_0^{t_0}\,dt = \int_0^\infty {dz\over
(1+z)H(z;b,w_0)},
\end{equation}
Similarly to the effect of bending parameter on the comoving
distance, in the case of age of the universe, increasing the bending
parameter makes a shorter age for the universe. Figure~\ref{fig:1}
shows the effect of bending parameter on the age of universe. Here
we show the variation of $H_0t_0$ as a function of $b$ for a typical
values of cosmological parameters (e.g. $h=0.65$, $\Omega_m=0.27$
and $w_0=-1.0$).

\begin{figure}
\epsfxsize=8.75truecm\epsfbox{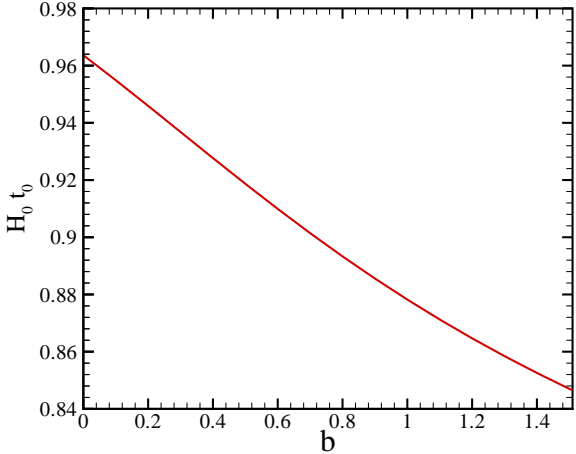} \narrowtext \caption{Age
times Hubble constant as the present time $(H_0t_0)$, as a
function of $b$ in a flat universe with the parameters of
$\Omega_{m}=0.3$, $h=0.65$ and $w_0=-1.0$. Increasing bending
parameters makes a shorter age for the universe. } \label{fig:1}
 \end{figure}

\end{itemize}

\section{Test by Supernova Type Ia: Gold Sample} \label{sn}
The Supernova Type Ia experiments provided the main evidence for
the existence of dark energy in the framework of standard
cosmology. Since 1995 two teams of the {\it High-Z Supernova
Search} and the {\it Supernova Cosmology Project} have been
discovered several type Ia supernova candidates at the high
redshifts \cite{per99,Schmidt}. Recently Riess et al. ~(2004)
announced the discovery of $16$ type Ia supernova with the Hubble
Space Telescope. This new sample includes $6$ of the $7$ most
distant ($z> 1.25$) type Ia supernovas. They determined the
luminosity distance of these supernovas and with the previously
reported algorithms, obtained a uniform Gold sample of type Ia
supernovas, containing $157$ objects \cite{R04,Tonry,bar04}. In
this section we compare the distance modulus of the Gold sample
data with that theoretically derived from the dark energy model.
The distance modulus $(\mu=m-M)$ in terms of redshift and
parameters of model is given by:
\begin{equation}
m-M=5\log{D_{L}(z;b,w_0)} + 25, \label{eq:mMr}
\end{equation}
where $M$ is the absolute magnitude, $D_{L}$ is the luminosity
distance in Mpc and $m$ is the corrected apparent magnitude,
including reddening, K correction etc. For a flat and homogeneous
cosmological model the luminosity distance can be obtained by:
\begin{eqnarray}\label{luminosity}
D_L (z;b,w_0) &=& (1+z)\int^{z}_{0}{dz'\over H(z';b,w_0)}.
\end{eqnarray}
The comparison between the observed and theoretical distance modulus
is given by $\chi^2$ as follows:
\begin{eqnarray}\label{chi_sn}
\chi^2=\sum_{i}^N\frac{[\mu_{obs}(z_i)-\mu_{th}(z_i;\Omega_m,w_0,b,h)]^2}{\sigma_i^2},
\end{eqnarray}
where $\sigma_i$ is the error bar of the observed distance modulus
for each Supernova candidate. The best fit values for the model
parameters are $w_0=-1.90_{-3.29}^{+0.75}$,
$\Omega_m=0.01^{+0.51}_{-0.01}$ and $b=6.00_{-6.00}^{+7.35}$ with
$\chi^2_{min}/N_{d.o.f} =1.13$ at $1 \sigma$ level of confidence.
Figure \ref{modul1} compares the distance modulus of the observed
SNIa Gold sample in terms of redshift and the best fit from the dark
energy model. We see clearly that the fit values in this model are
evidently different from those of $\Lambda$CDM (the WMAP results for
$\Lambda$CDM models are: $h=0.71^{+0.04}_{-0.03}$ and
$\Omega_m=0.27^{+0.04}_{-0.04}$ \cite{spe03,pearson03}). We also
compare our result with that of Riess et al. (2004), putting $b=0$
we should recover their result. Figure \ref{fig61} indicates the
confidence contours of $1\sigma$, $2\sigma$ and $3\sigma$ in the
$(\Omega_m,w_0)$ plane for the case of $b=0$, marginalized over $h$,
shows a good agreement with that of Riess et al. (2004).

By substituting the cosmological parameters derived from the SNIa fit
in equation (\ref{age}), we obtain the age of universe about $13.45$
Gry, which is in good agreement with the age of old stars.

\begin{figure}
\epsfxsize=9.75truecm
\epsfbox{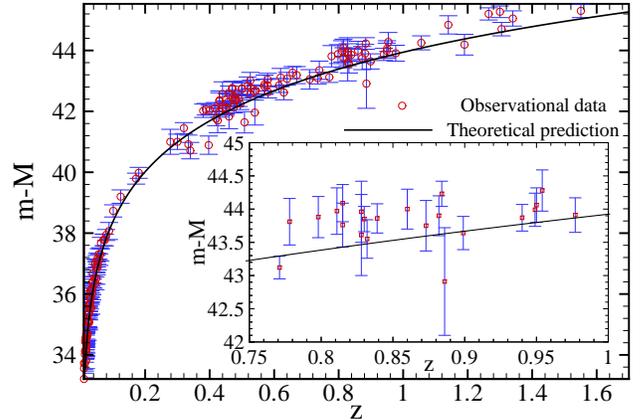} \narrowtext \caption{Comparison of the distance
modulus of the SNIa Gold sample in terms of redshift with the dark
energy model. Solid line shows the best fit ($\chi^2_{min}/N_{d.o.f}
=1.13$) with the corresponding parameters of
$w_0=-1.90_{-3.29}^{+0.75}$, $\Omega_m=0.01^{+0.51}_{-0.01}$ and
$b=6.00_{-6.00}^{+7.35}$ in $1 \sigma$ level of confidence. }
\label{modul1}
\end{figure}

\begin{figure}
\epsfxsize=8.75truecm\epsfbox{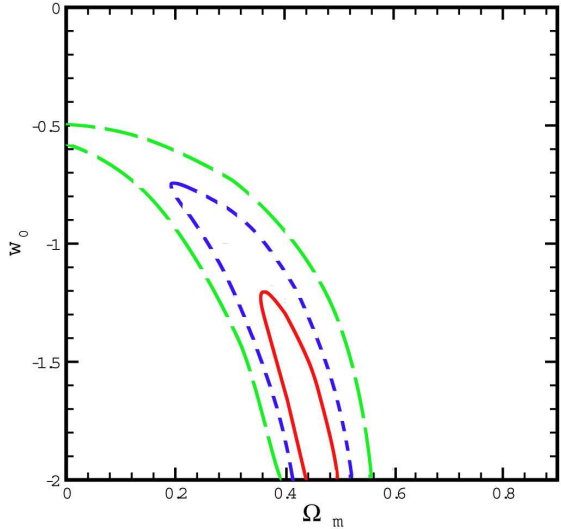} \narrowtext \caption {
Joint confidence intervals for $\Omega_m$ and $w_0$ for the case
of $b=0$ with  $1\sigma$ (solid-line), $2\sigma$ (dashed-line) and
$3\sigma$ (long dashed-line) confidence level. This result is in
good agreement with that of Riess et al. (2004).} \label{fig61}
\end{figure}

\section{The location of acoustic peak on CMB Map}
\label{cmb} In this section we use the CMB data from the WMAP
experiment to put additional constraints on the parameters of the
dark energy model \cite{b03}. The statistical properties of the
temperature fluctuations on CMB is given by a two point
correlation function. In the isotropic universe, the two point
correlation function $C(\gamma)$ depends only on the angle between
the two vectors $(\gamma)$ connecting the observer to the last
scattering surface and it can be expanded into a Legendre
polynomials as:
\begin{equation} C(\gamma) \,=\,
\frac{1}{4\pi}\sum_{l=2}^{\infty}(2l+1) C_l P_l(\cos{\gamma}),
\end{equation}
where $C_l$ is the widely used {\it angular power spectrum}.  The
summation over $l$ starts from $2$ because the $l\,=\,0$ term is the
monopole which in the case of statistical isotropy the monopole is
constant, and it can be subtracted out. The dipole $l\,=\,1$ is due
to the local motion of the observer with respect to the last
scattering surface and can be subtracted out as well.

The relevant parameter in the spectrum of CMB which can determine
the geometry and matter content of universe is the position of the
apparent acoustic peak. The acoustic peak corresponds to the Jeans
length of photon-baryon structures at the last scattering surface
some $\sim 379$ Kyr after the Big Bang \cite{spe03}. The position
of the acoustic peak in  Legendre polynomial space relates to its
apparent angular size, $\theta_A$ through $l_A\equiv
\pi/\theta_A$. The apparent angular size of acoustic peak in a
flat universe can be obtained by dividing of the comoving sound
horizon at the decoupling epoch $r_s(z_{dec})$ to the comoving
distance of observer to the last scattering surface $r(z_{dec})$
as:
\begin{equation}
\theta_A = {{r_s(z_{dec})}\over r(z_{dec}) }, \label{eq:theta_s}
\end{equation}
where we take the redshift of the decoupling at $z_{dec} = 1089$
\cite{Hu95}. The sound horizon corresponds to a distance that a
perturbation of pressure can travel from the beginning of universe
up to the decoupling area. The size of sound horizon at numerator
of equation (\ref{eq:theta_s}) can be obtained by:
\begin{equation}
r_s(z_{dec};b,w_0) =
\int_{z_{dec}}^{\infty}\frac{v_s(z)}{H(z;b,w_0)}dz \label{sh}
\end{equation}
where $v_s(z)^{-2}=3 + 9/4\times\rho_b(z)/\rho_r(z)$ is the sound
velocity in the unit of speed of light from the big bang up to the
last scattering surface \cite{dor01,Hu95}. The denominator of
equation (\ref{eq:theta_s}), $r(z_{dec})$, comoving distance to
the last scattering surface is also given by equation
(\ref{comoving}).

Changing parameters of the dark energy can shift the size of
apparent acoustic peak and the position of $l_A$ in the power
spectrum. Here we plot the dependence of $l_A$ on $b$ and $w_0$
for a typical values of cosmological parameters (Figure
\ref{lab}). It is seen that increasing $b$ makes a shorter
comoving distance to the last scattering surface (see Figure
\ref{fig:rz}) and subsequently results in a larger acoustic size
or smaller $l_A$. By a similar argument as in the case of $b$,
shifting $w_0$ towards zero makes smaller $l_A$.

\begin{figure}
\begin{center}
\epsfxsize=9.0truecm\epsfbox{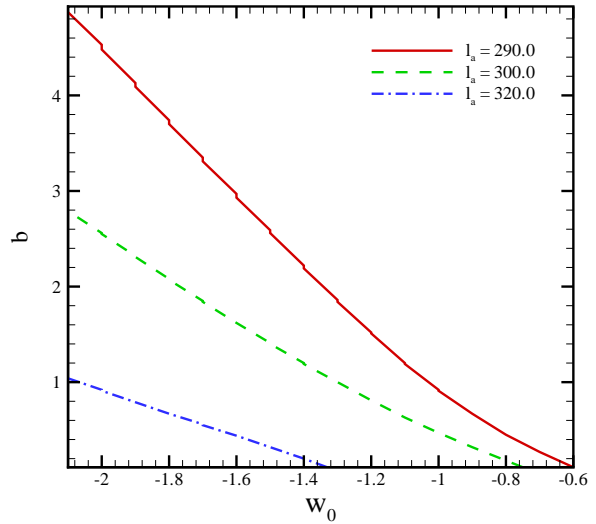} \narrowtext
\caption{Dependence of acoustic angular scale $l_A$ on the bending
parameter, $b$ and $w_0$ for the three
cases of $l_A=290$ (solid-line), $300$ (dashed-line) and $320$ (dashed-dotted line).}
\label{lab}
\end{center}
\end{figure}
In order to compare the angular size of acoustic peak from this
model with the observation we use the shift parameter as
\cite{bond97}:
\begin{equation}
\label{shift} R=
\sqrt{\Omega_m}\int_0^{z_{dec}}\frac{dz}{E(z;b,w_0)},
\end{equation}
where $E(z;b,w_0)=H(z;b,w_0)/H_0$. The shift parameter is
proportional to the size of acoustic peak to that of flat pure-CDM,
$\Lambda=0$ model, ($R\propto \theta_{A}/\theta_{A}^{\it
flat}=l_{A}^{\it flat}/l_{A}$). The observational result of CMB
experiments correspond a shift parameter of $R=1.716\pm0.062$ (given
by WMAP, CBI, ACBAR) \cite{spe03,pearson03}. One of the advantages
of using the parameter $R$ is that it is independent of the Hubble
constant. The best fit values for the dark energy model using the
combined CMB and SNIa observations, minimizing $\chi^2 =
\chi^2_{{\rm SNIa}} + \chi^2_{{\rm CMB}}$, results in: $\Omega_m =
0.37^{+0.08}_{-0.07}$, $b =2.7^{+5.25}_{-1.65}$ and $w_0 = -2.5^{+
0.90}_{-2.85}$ with $\chi_{min}/N_{d.o.f} = 1.11$ with ($1\sigma$)
level of confidence (see Table \ref{tab4}). The age of universe from
these parameters is $13.47$ Gyr.

In the next section we use the additional LSS data from SDSS and
combine them with that of CMB and SNIa, to put more constraint on
the parameters of the dark energy model.

\section{LSS Combined analysis with the CMB and SNIa}
\label{comb} Recent observations of large scale correlation
function measured from the spectroscopic sample of 46,748 {\it
Luminous Red Galaxies} (LRG) by the Sloan Digital Sky Survey shows
a well detected peak around $100$ Mpc $h^{-1}$.

This peak is an excellent match with the predicted shape and the
location of the imprint of the recombination-epoch acoustic
oscillation on the low-redshift clustering matter
\cite{eisenstein05}. For a flat universe we can construct
parameter $A$ as follows:
\begin{equation}
\label{lss1} A = \sqrt{\Omega_m}E(z_1;b,w_0)^{-1/3}
\times\left[\frac{1}{z_1}\int_0^{z_1}\frac{dz}{E(z;b,w_0)}\right]^{2/3},
\end{equation}
where $E(z;b,w_0) = H(z;b,w_0)/H_0$. Here $A$ is also independent
of $H_0$. We use the robust constraint on the dark energy model
using the value of $A=0.469\pm0.017$ from the LRG observation at
$z_1 = 0.35$ \cite{eisenstein05}.

In what follows we perform a combined analysis of SNIa, CMB and SDSS
to constrain the parameters of dark energy model. The combined
$\chi^2$ from each experiment is as follows:
\begin{equation}
\chi^2=\chi^2_{\rm {SNIa}}+\chi^2_{{\rm CMB}}+\chi^2_{{\rm SDSS}},
\end{equation}
where $\chi^2_{{\rm SNIa}}$ is obtained from equation
(\ref{chi_sn}), comparing the distance modulus of the SNIa
candidates from the Gold sample with that of theoretical dark energy
model. $\chi^2_{{\rm CMB}}$ is also calculated by comparison of the
observed shift parameter by WMAP with the model through equation
(\ref{shift}) and finally $\chi^2_{{\rm SDSS}}$ is obtained from
equation (\ref{lss1}). The marginalized likelihood functions
$(\mathcal{L}$$\propto e^{-\chi^2/2})$ \cite{press94} in terms of
the parameters of the model for three cases of (i) fitting model
with Supernova data, (ii) with SNIa$+$CMB data and (iii) considering
all three experiments of CMB$+$SNIa$+$SDSS are shown in Figure
\ref{mbhw}.

The best fit values for the model parameters by marginalizing on the
remained ones are: $\Omega_m=0.27_{-0.02}^{+0.02}$,
$b=1.35_{-0.90}^{+1.65}$ and $w_0=-1.45_{-0.60}^{+0.35}$ at
$1\sigma$ confidence level with $\chi^2_{min}/N_{d.o.f}=1.11$. Table
\ref{tab4} indicates the corresponding values for the cosmological
parameters from this fitting with one and two $\sigma$ level of
confidence. The joint confidence contours in the $(w_0,\Omega_m)$,
$(w_0,b)$ and $(b,\Omega_m)$ planes are shown in Figures \ref{jmw},
\ref{jbw} and \ref{jmb}. Comparing Figure \ref{fig61} which
corresponds to the fitting with SNIa, prior $b=0$ with the solid
contour in Figure \ref{jmw}, shows that adding the $b$ parameter
increases the degeneracy between $\Omega_m$ and $w_0$.

\begin{table}
\begin{center}
\caption{\label{tab4} The best values for the parameters of variable
dark energy as $\Omega_m$, $b$, $w_0$ with the corresponding
age for the universe from the fitting with the SNIa, SNIa+CMB and
SNIa+CMB+SDSS experiments with $1\sigma$ and $2\sigma$
confidence level.}
\begin{tabular}{|c|c|c|c|c|}
    Observation & $\Omega_m$& $b$& $w_0$& age \\
    &&&&(Gyr) \\ \hline
  &&&&\\
 & $0.01^{+0.51}_{-0.01}$&$6.00^{+7.35}_{-6.00}$ &$-1.90^{+0.75}_{-3.29}$ &  \\ 
 SNIa&&&&$13.45$\\
 & $0.01^{+0.56}_{-0.01}$&$6.00^{+17.42}_{-6.00}$ & $-1.90^{+1.10}_{-7.23}$ & \\
&&&&\\\hline
&&&&\\
    & $0.37^{+0.08}_{-0.07}$&$2.70^{+5.25}_{-1.65}$&  $-2.50^{+0.90}_{-2.85}$&  \\
 SNIa$+$CMB&&&& $13.47$\\

   &$0.37^{+0.18}_{-0.15}$&$2.70^{+11.85}_{-2.70}$&  $-2.50^{+1.50}_{-6.71}$&
\\
&&&&\\\hline
&&&&\\
SNIa$+$CMB& $0.27^{+0.02}_{-0.02}$&$1.35^{+1.65}_{-0.90}$ & $-1.45^{+0.35}_{-0.60}$&  \\
$+$SDSS &&&& $14.09$\\
&$0.27^{+0.04}_{-0.03}$&$1.35^{+6.30}_{-1.35}$  & $-1.45^{+0.65}_{-2.10}$& \\
&&&&\\
 \end{tabular}
\end{center}
\end{table}

We repeat the same analysis for the special case, fixing $w_0=-1.0$.
The best fit values for the parameters of the model in this case
obtain as $\Omega_m=0.28_{-0.02}^{+0.02}$, $b=0.24_{-0.23}^{+0.27}$
at $1\sigma$ confidence level with $\chi^2_{min}/N_{d.o.f}=1.12$.
For the case of using the result of HST-Key project (Hubble
parameter $h=0.71\pm0.07$), the marginalized likelihood function in
this special case, where we have two free parameters of $\Omega_m$
and $b$ are shown in Figure \ref{mb_hst}. The best fit values for
the model parameters are: $\Omega_m=0.20_{-0.01}^{+0.01}$,
$b=0.00_{-0.00}^{+0.02}$ at $1\sigma$ confidence level with
$\chi^2_{min}/N_{d.o.f}=1.75$. In this case we have almost
$\Lambda$CDM model with no variation in the equation of state of
dark energy.

Finally we do the consistency test, comparing the age of universe
derived from this model with the age of old stars and the age of
Old High Redshift Galaxies (OHRG) in various redshifts. One of the
reasons for the existence of the dark energy is the problem of
"age crisis", i.e. the universe without cosmological constant is
younger than its constituents \cite{dunlop96}. From Table
\ref{tab4}, we see that the age of universe from the combined
analysis of SNIa$+$CMB$+$SDSS is $14.09$ Gyr which is in agreement
with the age of old stars \cite{carretta00}. Here we take three
OHRG for comparison with the dark energy model, namely the LBDS
$53$W$091$, a $3.5$-Gyr-old radio galaxy at $z=1.55$
\cite{dunlop96}, the LBDS $53$W$069$, a $4.0$-Gyr-old radio galaxy
at $z=1.43$ \cite{dunlop99} and a quasar, APM $08279+5255$ at
$z=3.91$ with an age of $t=2.1_{-0.1}^{+0.9}$Gyr
\cite{hasinger02}. The later one has once again led to the "age
crisis". An interesting point about this quasar is that it cannot
be accommodated in the $\Lambda$CDM model \cite{jan06}. To
quantify the age consistency test we introduce the expression
$\tau$ as:
\begin{equation}
 \tau=\frac{t(z;b,w_0)}{t_{obs}} = \frac{t(z;b,w_0)H_0}{t_{obs}H_0},
\end{equation}
where $t(z)$ is the age of universe which can be obtained from the
equation (\ref{age}) and $t_{obs}$ is an estimation for the age of
old cosmological object. In order to have a compatible age for the
objects, we should have $\tau>1$.

Table \ref{tab6} shows the value of $\tau$ for three mentioned
OHRG. Various observational constraints on the parameters of the
dark energy model from SNIa, CMB, LSS and their combinations,
results an age for the universe more than the age of LBDS
$53$W$069$ and LBDS $53$W$091$, while APM $08279+5255$ at $z=3.91$
is older than the age of universe. Only in the case that we fix
$w_0=-1$ and use the SNIa+HST constraints, we obtain $\tau=1.26$,
a compatible age for the universe with that of Quasar.

\begin{table}
\begin{center}
\caption{\label{tab6} The value of $\tau$ for three high redshift
objects, using the parameters of the dark energy model from the
best fit.}
\begin{tabular}{|c|c|c|c|}
  Observation & LBDS $53$W$069$&LBDS $53$W$091$& APM  \\
&&& $08279+5255$ \\
  & $z=1.43$&$z=1.55$& $z=3.91$  \\ \hline
&&&\\
SNIa& $1.13$ & $1.21$& $0.78$ \\
, $(w_0=-1)$&&& \\
&&&\\\hline
&&&\\
 SNIa+HST & $1.67$&$1.81$&$1.26$ \\
, $(w_0=-1)$&&& \\
&&&\\ \hline

 SNIa$+$CMB & && \\
 $+$SDSS&$1.18$&$1.27$&$0.82$ \\
, $(w_0=-1)$& && \\ \hline
SNIa$+$CMB & && \\
 $+$SDSS&$1.29$&$1.38$&$0.89$ \\
$+$HST,& && \\ $(w_0=-1)$& && \\
\hline  & && \\ 
SNIa&$1.00$&$1.05$&$0.65$\\
&&&\\ \hline
SNIa$+$CMB & $1.09$&$1.17$&$0.75$ \\
 $+$SDSS& && \\
\end{tabular}
\end{center}
\end{table}
\begin{figure}
\epsfxsize=9.5truecm\epsfbox{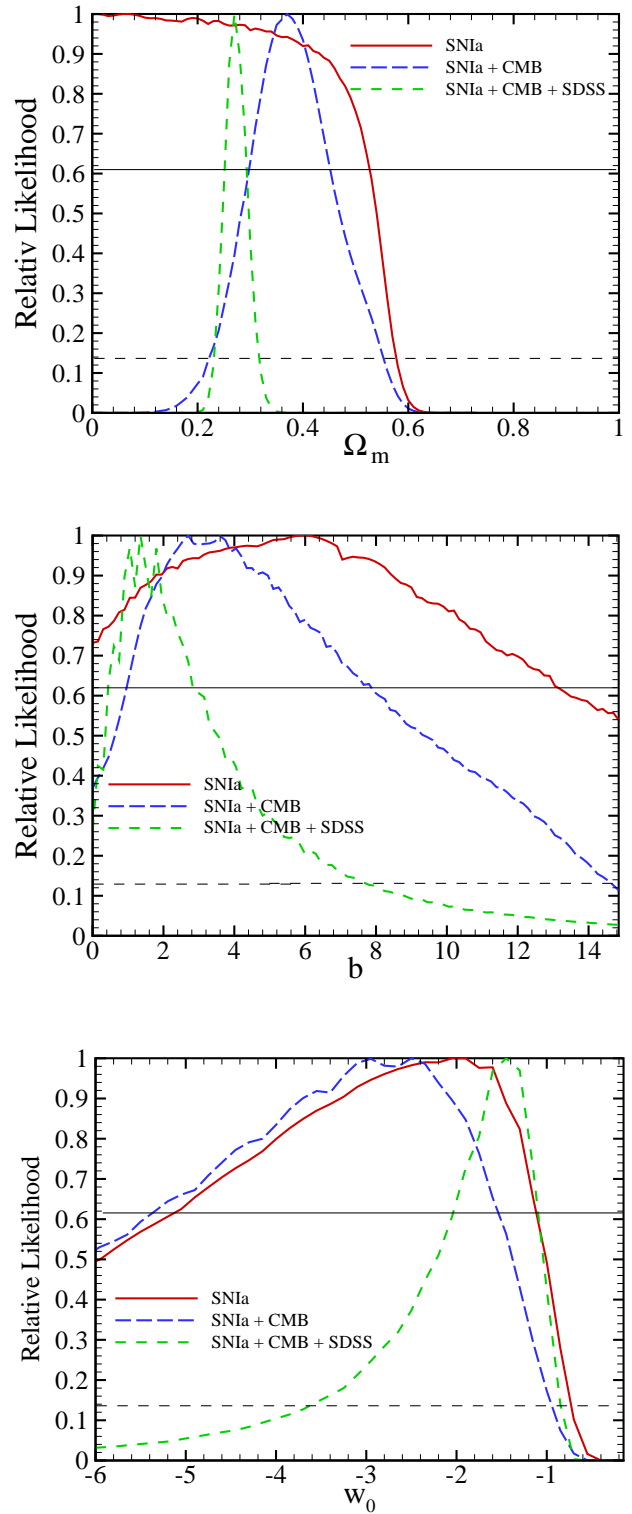} \narrowtext
\caption{Marginalized likelihood functions of three cosmological
parameters. The solid line is the likelihood function fitting the
model with SNIa data, the long dashed-line with the joint
SNIa$+$CMB and the dashed-line corresponds to the fitting with
SNIa$+$CMB$+$SDSS data. The intersections of the curves with the
horizontal solid and dashed lines give the bounds with $1\sigma$
and $2\sigma$ level of confidence respectively.} \label{mbhw}
 \end{figure}
\begin{figure}
\epsfxsize=9.5truecm\epsfbox{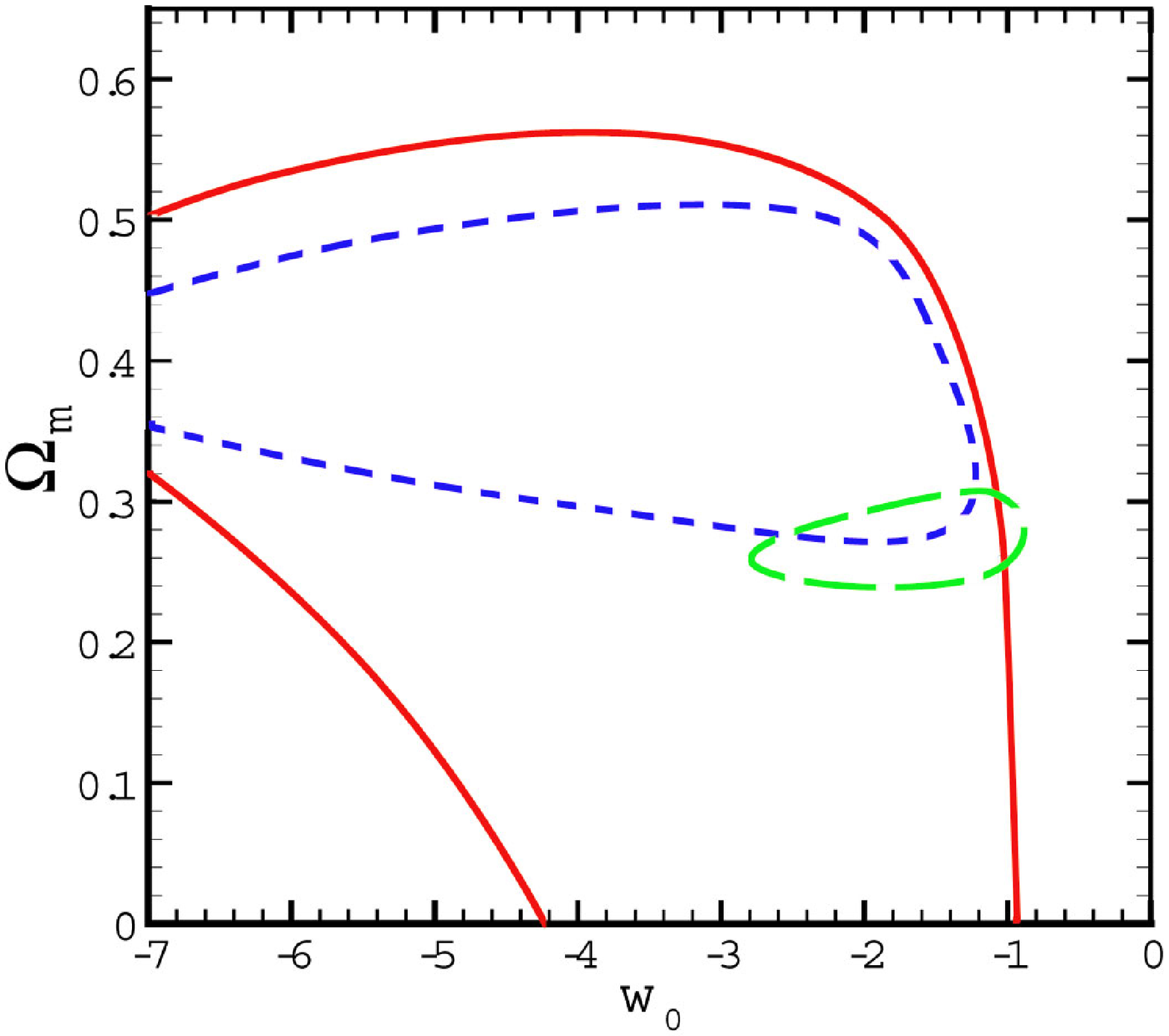} \narrowtext \caption{Joint
confidence intervals for $\Omega_m$ and $w_0$ from fitting with
the SNIa (solid line), SNIa$+$CMB (dashed-line) and
SNIa$+$CMB$+$SDSS (long dashed-line) data with $1\sigma$ level of
confidence.} \label{jmw}
 \end{figure}

\begin{figure}
\epsfxsize=9.5truecm\epsfbox{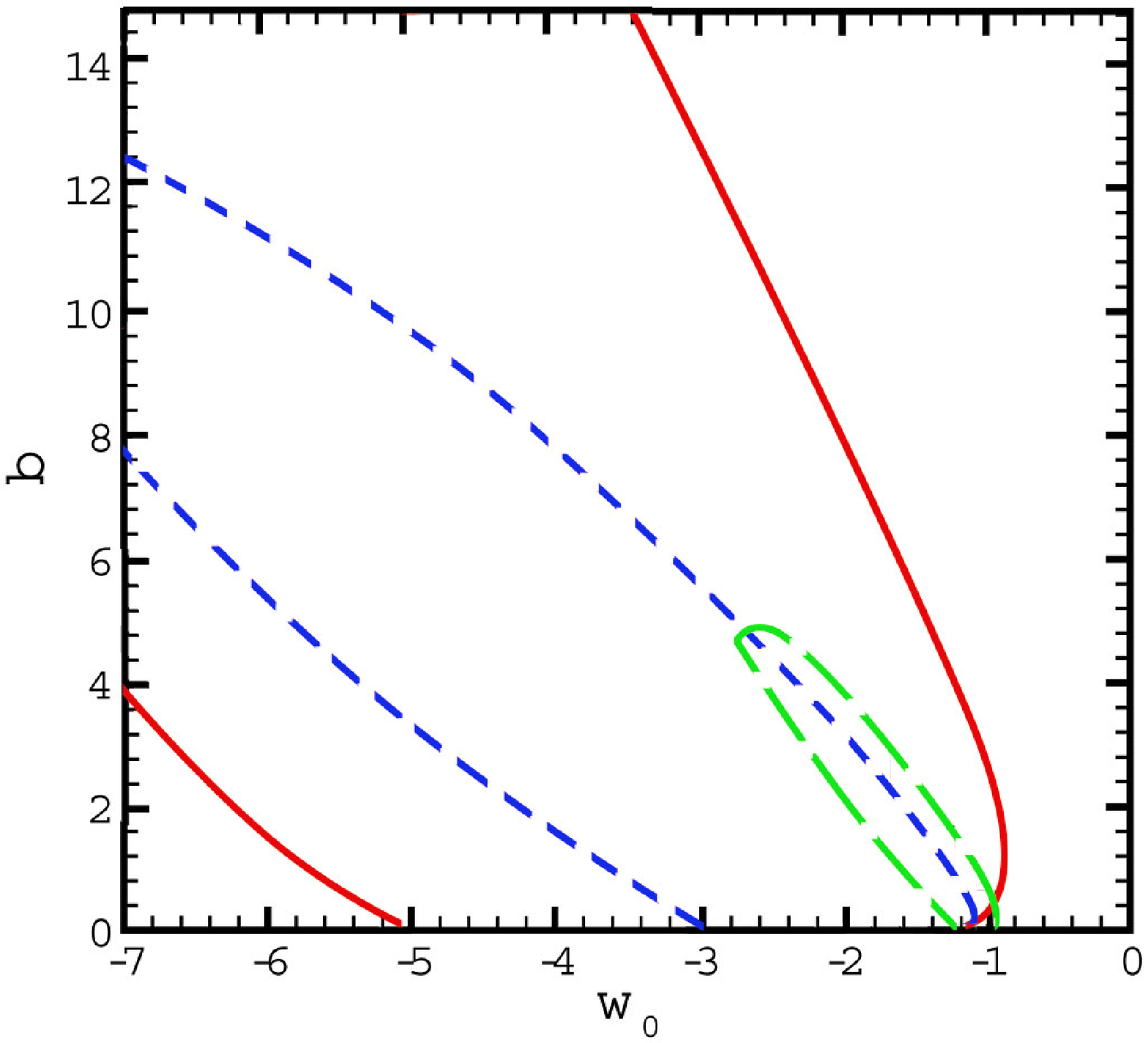} \narrowtext \caption{Joint
confidence intervals for $b$ and $w_0$ from fitting with the SNIa
(solid line), SNIa$+$CMB (dashed-line) and SNIa$+$CMB$+$SDSS (long
dashed-line) data with $1\sigma$ level of confidence.} \label{jbw}
 \end{figure}
\begin{figure}
\epsfxsize=9.5truecm\epsfbox{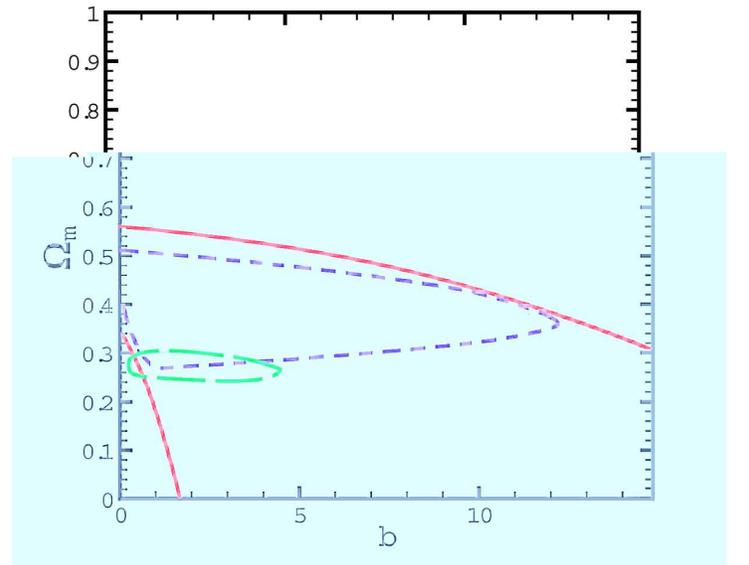} \narrowtext \caption{Joint
confidence intervals for $\Omega_m$ and $b$ from fitting with the
SNIa (solid line), SNIa$+$CMB (dashed-line) and SNIa$+$CMB$+$SDSS
(long dashed-line) data with $1\sigma$ level of confidence.}
\label{jmb}
 \end{figure}

\section{Conclusion}
\label{conc} In this work we examined a parameterized quintessence
model proposed by Wetterich (2004)\cite{w04} in a flat universe
with the low, intermediate and high redshift observations. The
effect of this model on the age of universe, radial comoving
distance, comoving volume element and the variation of apparent
size of objects with the redshift (Alcock-Paczynski test) have
been studied.

In order to constrain the parameters of model we used the Gold
sample SNIa data. The supernova analysis results in a large
degeneracy between the parameters of the model. To improve the
analysis, we combined analysis of SNIa and CMB shift parameter.
Finally we used the results of recently observed baryonic peak at
the Large Scale Structure, combined with the two former experiments
to confine the parameters of the dark energy model. The fitting
parameters from the joint analysis of SNIa$+$CMB $+$ SDSS
marginalizing on the remained ones results in:
$\Omega_m=0.27_{-0.02}^{+0.02}$, $b=1.35_{-0.90}^{+1.65}$ and
$w_0=-1.45_{-0.60}^{+0.35}$ at $1\sigma$ confidence level with
$\chi^2_{min}/N_{d.o.f}=1.11$. The best fit value for the equation
of state of the dark energy leads to $w_0<-1$ which violates the
strong energy condition. From the quantum field theory point of
view, exotic models like scalar field with negative kinetic energy
can provide $w<-1$ \cite{car03,ha73} (theoretical attempts for $w
<-1$ can be found in
~\cite{Caldwell:1999ew,parker,frampton,Ahmed:2002mj,CHT})

\begin{figure}
\epsfxsize=9truecm\epsfbox{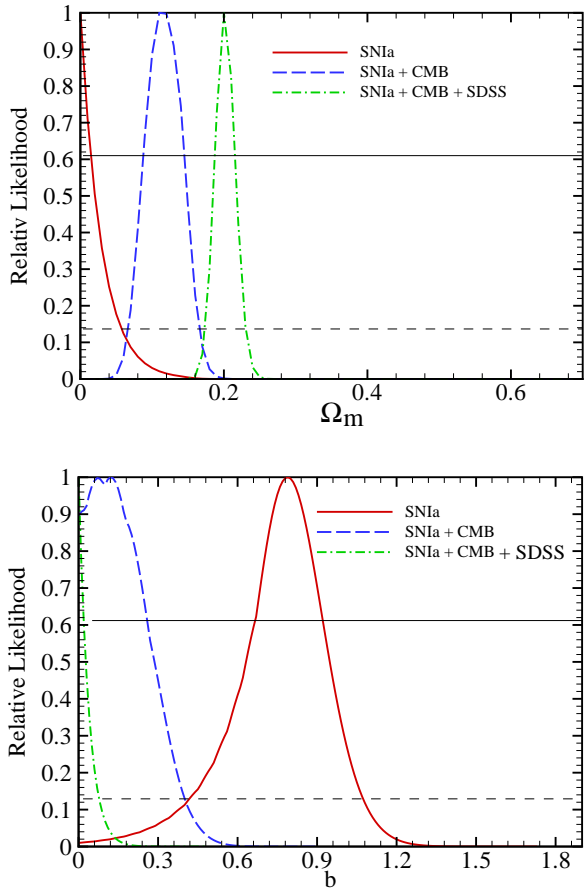} \narrowtext
\caption{Marginalized likelihood functions for two cosmological
parameters of $\Omega_m$ and $b$ in the special case of fixing
$w_0=-1$ and using the $H_0=71.0\pm7.0$ from the HST-Key
project.The solid line corresponds to the Likelihood function of
the parameters for the case of fitting dark energy model with the
SNIa data, the dashed line corresponds to the combined SNIa$+$
CMB data and dotted-dashed line is for combined SNIa$+$CMB$+$SDSS
data. The intersections with the horizontal solid and dashed
lines give the bounds for $1\sigma$ and $2\sigma$ confidence
level respectively.} \label{mb_hst}
\end{figure}
We also did the age test, comparing the age of old stars and old
high redshift galaxies with the age that we obtained based on the
dark energy model. The age of the universe from the best fit
parameters of the model, results in an age of $14.09$ Gyr for the
universe which is in agreement with the age of old stars. We also
chose two high redshift radio galaxies at $z=1.55$ and $z=1.43$
with a quasar at $z=3.91$. The two first objects were consistent
with the age of the universe, meaning that there were younger than
the age of universe at the corresponding redshifts while the
latter one was older than the age of universe. The age of APM
$08279+5255$ quasar as the "age crisis" was not compatible with
the age of universe in this dark energy model. Only in the case
that we fixed $w_0=-1$ and took $H_0=71.0\pm 7.0$ a compatible age
for the universe with that of quasar has been obtained.


{\bf Acknowledgements}
The authors thank the anonymous referee for useful comments.
This paper is dedicated to Dr. Somaieh Abdolahi.

\end{document}